# Theory of Phonon Shakeup Effects on Photoluminescence from the Wigner Crystal in a Strong Magnetic Field

D. Z. Liu[1*], H.A. Fertig[2], and S. Das Sarma[1]

[1] *Center for Superconductivity Research, Department of Physics, University of Maryland, College Park, Maryland 20742*

[2] *Department of Physics and Astronomy, University of Kentucky, Lexington, Kentucky 40506-0055*

We develop a method to compute shakeup effects on photoluminescence from a strong magnetic field induced two-dimensional Wigner crystal. Only localized holes are considered. Our method treats the lattice electrons and the tunneling electron on an equal footing, and uses a quantum-mechanical calculation of the collective modes that does not depend in any way on a harmonic approximation. We find that shakeup produces a series of sidebands that may be identified with maxima in the collective mode density of states, and definitively distinguishes the crystal state from a liquid state in the absence of electron-hole interaction. In the presence of electron-hole interaction, sidebands also appear in the liquid state coming from short-range density fluctuations around the hole. However, the sidebands in the liquid state and the crystal state have different qualitative behaviors. We also find a shift in the main luminescence peak, that is associated with lattice relaxation in the vicinity of a vacancy. The relationship of the shakeup spectrum with previous mean-field calculations is discussed.

## I. INTRODUCTION

It was first pointed out by Wigner[1] more than sixty years ago that an electron gas will undergo a zero-temperature, quantum phase transition into a crystalline phase as the density is lowered, since quantum fluctuation effects diminish more rapidly than Coulomb correlation. Only two decades ago, the first convincing evidence of an electron crystal was presented for a system of electrons on a He surface[2]. The electron densities attainable in this fashion are extremely low, however, making this an unattractive system for observing the quantum phase transition. Semiconductors are much more attractive systems in this sense, because one has great control over the electron densities, through dopant concentrations. A particularly good candidate for observing the Wigner crystal (WC) is the two-dimensional electron gas (2DEG), as realized in modulation doped semiconductors. Samples of this type are now available with such high quality that the electron groundstate is not necessarily dominated by disorder. The possibility of observing the WC is further enhanced by the application of a strong perpendicular magnetic field, which quenches the kinetic energy, and allows the formation of a crystal state at higher densities (for which disorder effects are less important) than would be possible without it.

Experimental evidence which may be associated with the WC in 2DEG's has accumulated over the last several years[3], including rf data, transport experiment, cyclotron resonance, and photoluminescence (PL) experiments. In the last of these probes, which has produced much intriguing data, either a valence band hole[4] or a hole bound to an acceptor[5,6] recombines with an electron in the 2DEG, producing a characteristic photon spectrum. A mean-field analysis[7] of the latter type of experiment showed that the PL spectrum has, in principle, characteristic signatures of the WC: a "Hofstadter butterfly"[8] spectrum for the case of weak interactions between the electrons and the hole, and a characteristic shift in the PL spectrum upon melting of the crystal.

In this paper, we go beyond the mean-field approximation, to examine shakeup effects on the PL spectrum; i.e., we will examine how the collective mode spectrum of the WC (which, at long wavelengths, corresponds to the classical phonon spectrum), and the fact that some of these modes may be excited in the electron-hole recombination process, modify the results of the mean-field theory. We will consider in detail only the case of a localized hole[5,6]. Our method is purely quantum-mechanical, and treats both the tunneling electron and the other lattice electrons on the same footing. Furthermore, we employ a quantum treatment of the collective modes to realistically account for contributions both from small and large wavevector collective excitations of the lattice. This approach has the further advantage of depending in no way upon a harmonic approximation for the lattice[9]. The method allows in principle for shakeup of arbitrary numbers of these excitations. Since we are working in the strong magnetic field limit, we consider only excitations within the lowest Landau level (LLL). Our main results are: (1) Shakeup effects shift the main PL peak to higher energies than found in a mean-field treatment[7]. (2) The Hofstadter spectrum is eliminated from the PL spectrum, even in the case of weak electron-hole interactions[6] (although we will argue below that it survives in the itinerant hole case[7]). The sudden shift of the PL spectrum upon melting, by contrast, survives even when shakeup is included. (3) Phonon sidebands appear that corre-spond to maxima in the phonon density of states (DOS); some (but not all) of these sidebands are results of van Hove singularities in the DOS, and so are characteristic of an ordered WC state. For the case of weak electron-hole interactions[6], we *do not* see these sidebands in the liquid state, so that phonon satellites uniquely distinguish between a liquid and a solid state. Interestingly, for stronger electron-hole interactions, a shakeup satellite persists even above the melting temperature. We expect this sideband to lose oscillator strength relative to the main peak, either with increasing temperature or decreasing electron-hole interaction strength. The latter may be accomplished by examining PL from several samples with different acceptor - 2DEG setback distances[6].

This paper is organized in the following way. In Sec II, we review the mean-field theory for the photoluminescence, as well as the time-dependent Hartree-Fock approximation for collective modes in strong magnetic fields, and show how they may be combined to calculate the phonon shake-up effect on photoluminescence. We present and discuss our numerical results in Sec. III, and make some concluding remarks in Sec.IV. A brief account of some of these results has been published previously.[10]

## II. THEORY

In this section, we first set up the theoretical framework by reviewing the general expressions for the time-dependent Hartree-Fock approximation(TDHFA) of the 2DEG in a strong magnetic field and outlining the mean-field theory for photoluminescence from Wigner crystal. Then we introduce the expressions for collective modes in Wigner crystal. Finally we present the detailed formalism to calculate the collective modes shake-up effects on the photoluminescence.

### A. Single-Particle Properties

It is well known that non-interacting two-dimensional electrons in a perpendicular magnetic field ($\mathbf{B} = -B\hat{z}$) have an energy spectrum of discrete Landau levels: $E_N = (N + \frac{1}{2})\hbar\omega_c$, $N = 0, 1, 2, ...$, where $\omega_c = eB/m^*c$ is the cyclotron resonance frequency. Working in the Landau gauge($\mathbf{A} = -Bx\hat{y}$), and with periodic boundary conditions in the $\hat{y}$-direction, the single particle eigenstates are given by:

$$<\mathbf{r}|NX> = \frac{1}{L_y} exp(iXy/l_c^2)\phi_N(x-X). \quad (1)$$

Here $l_c = (\hbar c/eB)^{1/2}$ is the magnetic length and $\phi_N$ is the one-dimensional harmonic oscillator eigenstate with oscillation centers $X$. The allowed values of $X$ are separated by $2\pi l_c^2/L_y$. The degeneracy of each Landau level is given by $g = S/2\pi l_c^2$, with $S$ the area of the 2DEG.

To address the problem of photoluminescence from a strong magnetic field induced two-dimensional WC, we now consider a system with an interacting 2DEG and a layer with a low density of holes (either localized or itinerant) separated from the electron plane by a distance $d$, in the presence of a strong perpendicular magnetic field. We assume the magnetic field is very strong; *i.e.*, $\hbar\omega_c >> e^2/\epsilon a$, and the electronic filling factor $\nu = nhc/eB << 1$ in the WC regime, so that we can assume only the lowest Landau level is occupied by electrons. The general many-body Hamiltonian can be expressed in terms of single-electron eigenstates (only $N = 0$ eigenstates are included) as follows:

$$\mathcal{H} = \mathcal{H}_o + \mathcal{H}_{ee} + \mathcal{H}_{eh}, \quad (2)$$

where

$$\mathcal{H}_o = \sum_X \frac{1}{2}\hbar\omega_c a_X^\dagger a_X + \sum_i E_h c_i^\dagger c_i$$

$$\mathcal{H}_{ee} = \frac{1}{2S}\sum_{\mathbf{q}\neq 0}\sum_{X_1 X_2 X_3 X_4} V_c(\mathbf{q}) <X_1|exp(i\mathbf{q}\cdot\mathbf{r})|X_4>$$
$$\times <X_2|exp(-i\mathbf{q}\cdot\mathbf{r})|X_3> a_{X_1}^\dagger a_{X_2}^\dagger a_{X_3} a_{X_4}$$

$$\mathcal{H}_{eh} = \frac{1}{S}\sum_{\mathbf{q}}\sum_{X_1 X_2}\sum_{ij} V_{eh}(\mathbf{q}) <X_1|exp(i\mathbf{q}\cdot\mathbf{r})|X_2>$$
$$\times <i|exp(-i\mathbf{q}\cdot\mathbf{r})|j> a_{X_1}^\dagger a_{X_2} c_i^\dagger c_j. \quad (3)$$

Here $\mathcal{H}_o$ is the single-particle zero-point energy which is constant. $\mathcal{H}_{ee}$ is the electron-electron interaction, in which $V_c(\mathbf{q}) = 2\pi e^2/\epsilon q$ is the two-dimensional Fourier transform of the Coulomb interaction. $\mathcal{H}_{eh}$ is the electron-hole interaction, in which $V_{eh}(\mathbf{q})$ is the Fourier transform of the interaction between one electron and one hole. For a system with a small density of holes, it is a good approximation to also ignore hole-hole interactions, which are extremely small in comparison with $\mathcal{H}_{eh}$. In our current consideration, we also do not include the effect of impurities and any external potential. In practice, we may also drop $\mathcal{H}_o$ because it is just a constant. The matrix elements in Eq.(3) are given by

$$<X_1|exp(i\mathbf{q}\cdot\mathbf{r})|X_2>$$
$$= exp(\frac{i}{2}q_x(X_1 + X_2) - \frac{q^2 l_c^2}{4})\delta_{X_1, X_2+q_y l_c^2} \quad (4)$$

It is well known that the ground state configuration for a two-dimensional Wigner crystal is a triangular lattice[11]. In the presence of electron-hole interaction, the Wigner lattice should deform around the position of the hole. In order to take into account this effect, while still taking advantage of the periodicity of the WC, we divide the WC system into hexagon unit cells. Each supercell contains a finite number of electrons, and one hole (in the center, for localized hole case). The finite size of our unit cells will not be significant for large enough supercells.

Denoting $b$ as the distance between the centers of nearest-neighbor supercells, then the basis vectors for the superlattice are $\mathbf{L}_1 = b(1,0,0)$ and $\mathbf{L}_2 = b(\frac{\sqrt{3}}{2}, \frac{1}{2}, 0)$. Then the basis vectors for the reciprocal lattice are given by $\mathbf{G}_1 = G_0(1,0,0)$ and $\mathbf{G}_2 = G_0(-\frac{1}{2}, \frac{\sqrt{3}}{2}, 0)$, where $G_0 = \left(\frac{\sqrt{3}}{2}b\right)^{-1}$.

In a lattice system, all the expectation values for single particle properties in momentum space are non-zero only for momenta equal to reciprocal lattice vectors. We first define the density operator:

$$n(\mathbf{G}) = \int d^2\mathbf{r} \, exp(-i\mathbf{G} \cdot \mathbf{r}) n(\mathbf{r}) = g\rho(\mathbf{G}) exp(-\frac{G^2 l_c^2}{4}), \tag{5}$$

where

$$\rho(\mathbf{G}) = \frac{1}{g} \sum_{X_1 X_2} e^{-\frac{i}{2}G_x(X_1+X_2)} \delta_{X_1, X_2 - G_y l_c^2} a^\dagger_{X_1} a_{X_2}. \tag{6}$$

One may easily show that $<\rho(\mathbf{G}=0)> = <\hat{N}_e>/g = \nu$, where $\hat{N}_e$ is the electron number operator.

In terms of many-body operator decomposition, Hartree-Fock approximation can be expressed as:

$$a^\dagger_{X_1} a^\dagger_{X_2} a_{X_3} a_{X_4} = <a^\dagger_{X_1} a_{X_4}> a^\dagger_{X_2} a_{X_3} - <a^\dagger_{X_1} a_{X_3}> a^\dagger_{X_2} a_{X_4} \tag{7}$$

In this effective mean-field theory, a single electron moves as if in an average external potential provide by other electrons through both the direct (first term) and exchange (second term) interactions. For localized holes which are very far away from each other, we can also apply the approximation that $<c^\dagger_i c_j> = \delta_{ij}$. After some algebra, Hartree-Fock Hamiltonian for the 2DEG (with an electron-hole interaction appropriate for a localized hole) in the lowest Landau level can be written as:

$$\mathcal{H}_{HF} = g \sum_{\mathbf{G}} \left[ W(\mathbf{G}) <\rho(\mathbf{G})> + n_h V_{eh}(\mathbf{G}) e^{-G^2 l_c^2/4} \right] \rho(\mathbf{G}) \tag{8}$$

where $n_h$ is the density of holes. Note that we ignore the negative sign in front of $\mathbf{G}$ because of apparent symmetry in this problem. $W(\mathbf{G})$ is the effective Hartree-Fock interaction:

$$W(\mathbf{G}) = \frac{e^2}{\epsilon l_c^2} \left[ \frac{1}{Gl_c} e^{-G^2 l_c^2/2} (1 - \delta_{\mathbf{G},0}) - \sqrt{\frac{\pi}{2}} e^{-G^2 l_c^2/4} I_0\left(\frac{-G^2 l_c^2}{4}\right) \right] \tag{9}$$

where $I_0(x)$ is the modified Bessel function of the first kind[12].

We define single electron Green's function:

$$G(X_1, X_2; \tau) = - <T_\tau a_{X_1}(\tau) a^\dagger_{X_2}(0)> \tag{10}$$

It is convenient to define the Fourier transform

$$G(\mathbf{G}, \tau) = \frac{1}{g} \sum_{X_1 X_2} e^{-\frac{i}{2}G_x(X_1+X_2)} \delta_{X_1, X_2 + G_y l_c^2} G(X_1, X_2; \tau). \tag{11}$$

We will use this form of the Fourier transform throughout this work. The TDHFA is derived by writing the equation of motion for the Green's function,

$$\frac{\partial}{\partial \tau} G(X_1, X_2; \tau) = -\frac{\partial}{\partial \tau} <T_\tau a_{X_1}(\tau) a^\dagger_{X_2}(0)> \tag{12}$$
$$= -\delta_{X_1 X_2} \delta(\tau) - <T_\tau [\mathcal{H} - \mu_e N_e, a_{X_1}](\tau) a^\dagger_{X_2}(0)>.$$

One can compute the commutators explicitly, and simplify the result using a Hartree-Fock decomposition presented in Eq.(7). After Fourier transforming with respect to time, the equation of motion for $G(\mathbf{G}, \omega_n)$ can be written as[12]

$$(i\omega_n + \mu_e) G(\mathbf{G}, i\omega_n) - \sum_{\mathbf{G}'} B(\mathbf{G}, \mathbf{G}') G(\mathbf{G}', i\omega_n) = \delta_{\mathbf{G},0}, \tag{13}$$

where

$$B(\mathbf{G}_1, \mathbf{G}_2) = [W(\mathbf{G}_1 - \mathbf{G}_2) < \rho(\mathbf{G}_1 - \mathbf{G}_2) >$$
$$+ n_h V_{eh}(\mathbf{G}_1 - \mathbf{G}_2) e^{-(\mathbf{G}_1 - \mathbf{G}_2)^2 l_c^2/4}] e^{i(\mathbf{G}_1 \times \mathbf{G}_2) l_c^2/2}. \tag{14}$$

We can directly diagonalize matrix $B$ and obtain its eigenvectors $V_j(\mathbf{G})$ and eigenvalues $\omega_j^e$, after which the Green's function can be written as

$$G(\mathbf{G}, i\omega_n) = \sum_j \frac{V_j(\mathbf{G}) V_j^*(\mathbf{G} = 0)}{i\omega_n + \mu_e - \omega_j^e}$$
$$= \sum_j \frac{W_e(\mathbf{G}, j)}{i\omega_n + \mu_e - \omega_j^e}. \tag{15}$$

The density of states for electrons is then given by:

$$D(E) = -\frac{1}{\pi} Im(G(\mathbf{G} = 0, E + i\delta))(\frac{g}{S})$$
$$= -\frac{1}{\pi} Im(\sum_j \frac{W_e(\mathbf{G} = 0, j)}{E - \omega_j^e + i\delta})(\frac{1}{2\pi l_c^2}) \tag{16}$$

Finally, the density operator can be expressed as

$$< \rho(\mathbf{G}) > = G(\mathbf{G}, \tau = 0^-)$$
$$= \sum_j V_j(\mathbf{G}) V_j^*(\mathbf{G} = 0) f_{FD}(\omega_j^e - \mu_e). \tag{17}$$

Here $f_{FD}(x) = [1 + exp(\beta x)]^{-1}$ is the Fermi-Dirac distribution. Since $< \rho(\mathbf{G} = 0) > = \nu$, we can self-consistently calculate the chemical potential $\mu_e$, the density of states and the electron density in $\mathbf{G}$ space. By iteratively solving Eqs.(13),(15), and (17), we can calculate the density configurations for the Wigner crystal.

### B. Mean-Field Theory for Localized Holes

We now present our theory for photoluminesence from the WC in a strong magnetic field. The photoluminescence intensity is given, for a single localized hole state, by

$$P(\omega) = \frac{I_0}{Z} \sum_n \sum_m e^{-E_n/k_B T} | < m, 0|\hat{L}|n, h > |^2 \delta(\omega - E_n + E_m), \tag{18}$$

where $Z = \Sigma_n e^{-E_n/k_B T}$, $|n, h >$ is a many-body electron state with energy $E_n$ and $N$ electrons when there is a core hole present, $|m, 0 >$ is a many-body electron state with $N - 1$ electrons and energy $E_m$, $\omega$ is the luminescence frequency, and $\hat{L} = \int d^2\mathbf{r} \psi(\mathbf{r}) \psi_h(\mathbf{r})$ is the luminescence operator, with $\psi(x)$ the electron annihilation operator and $\psi_h(x)$ the hole annihilation operator. As written, the initial state is actually higher in energy than the final state, and we find it convenient to rework the problem in terms of absorption rather than emission. To accomplish this, we add a term $H' = -E_0 c_0^\dagger c_0$ to the Hamiltonian, where $c_0^\dagger$ creates a localized hole, and take the limit $E_0 \to \infty$. It is not difficult to show

$$P(\omega) = \lim_{E_0 \to \infty} \frac{P'(\omega - E_0)}{n_0(E_0)} \tag{19}$$

where $P'$ is the absorption spectrum of the new Hamiltonian, and $n_0$ is the average occupation of the hole state, which just becomes one in the limit $E_0 \to \infty$. The absorption spectrum is identical to Eq.(18), except one needs to add the energy $E_0$ to all the quantities $E_n$ in the expression. After standard manipulations[13], one can show that

$$P'(\omega) = \frac{I_0}{\pi} \frac{1}{1 - e^{\omega/k_B T}} Im \ R(\omega + i\delta) \tag{20}$$

The function $R(\omega + i\delta)$ is a response function, which continued to imaginary frequency has the form

$$\mathcal{R}(i\omega_n) = -\int_0^\beta < T_\tau L(\tau) L^\dagger(0) > e^{i\omega_n \tau} d\tau \tag{21}$$

To compute this quantity, we consider (for the case of a localized hole state) instead of a single hole, a periodic (hexagonal) lattice of them, with a unit cell that contains as many electrons as can be handled numerically. We allow neither interactions between the holes nor tunneling between the hole sites, so that in the limit of large unit (super)cells, one should expect the result to be the same as for the isolated hole case. Expanding the electron and hole creation and annihilation operators in $\hat{L}$ in Eq.(21) in terms of electronic states in the lowest Landau level and localized hole states, one finds

$$R(\omega) = \frac{n_h S}{2\pi l_c^2} \sum_{\mathbf{G}} R(\mathbf{G}, \omega) e^{-G^2 l_c^2/4}, \tag{22}$$

where $R(\mathbf{G}, \omega)$ is the Fourier transform of the following quantity:

$$\mathcal{R}_{ij}(X_1, X_2; \tau) = - < T_\tau a_{X_1}(\tau) c_i(\tau) c_j^\dagger(0) a_{X_2}^\dagger(0) >, \tag{23}$$

and

$$R_{ij}(\mathbf{G}, \tau) = \frac{1}{g} \sum_{X_1 X_2} e^{-\frac{i}{2} G_x (X_1 + X_2)} \delta_{X_1, X_2 + G_y l_c^2} \mathcal{R}_{i,j}(X_1, X_2; \tau). \tag{24}$$

We write down the equation of motion for $\mathcal{R}_{ij}(X_1, X_2; \tau)$ in terms of its commutator with the Hamiltonian:

$$\begin{aligned}
\frac{\partial}{\partial \tau} \mathcal{R}_{ij}(X_1, X_2; \tau) &\equiv -\frac{\partial}{\partial \tau} < T_\tau a_{X_1}(\tau) c_i(\tau) c_j^\dagger(0) a_{X_2}^\dagger(0) > \\
&= - < [a_{X_1} c_i, c_j^\dagger a_{X_2}^\dagger] > \delta(\tau) - < T_\tau [\mathcal{H}_{eff} - \mu(N_e - N_h), a_{X_1} c_i](\tau) c_j^\dagger a_{X_2}^\dagger >,
\end{aligned} \tag{25}$$

where $\mathcal{H}_{eff} = \mathcal{H} - E_0 \sum_i c_i c_i^\dagger$ and $N_e$ and $N_h$ are corresponding electron and hole number operator. Following steps closely analogous to those used to find the equation of motion for the Green's function in Sec. A above, the Hartree-Fock approximation for Eq.(25) becomes[7,12]

$$-(\omega + i\delta) R_{ij}(\mathbf{G}, \omega) = <\rho(\mathbf{G})> \delta_{ij} + (E_0 - \frac{1}{2}\hbar\omega_c - E_h) R_{ij}(\mathbf{G}, \omega) \tag{26}$$

$$- \sum_{\mathbf{G}'} W(\mathbf{G}') <\rho(\mathbf{G}')> e^{-i(\mathbf{G} \times \mathbf{G}') l_c^2/2} R_{ij}(\mathbf{G} - \mathbf{G}', \omega)$$

$$- \frac{1}{2\pi l_c^2} \sum_{\mathbf{G}'} V_{eh}(-\mathbf{G}') <\rho(\mathbf{G}')> e^{-G'^2 l_c^2/4} R_{ij}(\mathbf{G}, \omega)$$

$$- n_h \sum_{\mathbf{G}'} V_{eh}(\mathbf{G}') e^{-G'^2 l_c^2/4 - i(\mathbf{G} \times \mathbf{G}') l_c^2/2} R_{ij}(\mathbf{G} - \mathbf{G}', \omega). \tag{27}$$

It is apparent that the solution to this satisfies $R_{ij}(\mathbf{G}, \omega) = R(\mathbf{G}, \omega) \delta_{ij}$. This result may be expressed in the form

$$\sum_{\mathbf{G}'} \left[ (\omega + i\delta + E_0 - \omega_0) \delta_{\mathbf{G}, \mathbf{G}'} - B(\mathbf{G}, \mathbf{G}') \right] R(\mathbf{G}', \omega) = - <\rho(\mathbf{G})>, \tag{28}$$

where

$$\omega_0 = \frac{1}{2}\hbar\omega_c + E_h + \frac{1}{2\pi l_c^2} \sum_{\mathbf{G}} <\rho(\mathbf{G})> V_{eh}(-\mathbf{G}) e^{-G^2 l_c^2/4}, \tag{29}$$

and $B$ is exactly given by Eq.(14).

We see that the form of $R$ is essentially that of electron Green's function as in Eq.(13). By inverting Eq.(28), we have

$$\begin{aligned}
R(\mathbf{G}, \omega) &= -\sum_{\mathbf{G}'} \left[ (\omega + i\delta + E_0 - \omega_0) \delta_{\mathbf{G}, \mathbf{G}'} - B(\mathbf{G}, \mathbf{G}') \right]^{-1} <\rho(\mathbf{G}')> \\
&= -\sum_{j\mathbf{G}'} \frac{V_j(\mathbf{G}) V_j^{-1}(\mathbf{G}') <\rho(\mathbf{G}')>}{\omega + i\delta - \omega_0 - \omega_j^e} \\
&= -\sum_j \frac{W_e(\mathbf{G}, j) f_{FD}(\omega_j^e - \mu_e)}{\omega + i\delta - \omega_0 - \omega_j^e}.
\end{aligned} \tag{30}$$

Here we have already dropped $E_0$ because it cancels out when we calculate the final photoluminescence power using Eqs.(19) and (20). We can see that $R$ has poles at precisely the same energies as the poles in the electron Green's function for the system in the presence of the external interaction $V_{eh}$ due to the hole[14], up to the constant energy shift $\omega_0$. However, the residues of the poles are not the same as for the Green's function. The residues are determined by the overlaps of the mean-field single particle electron wavefunctions with the hole wavefunction. It is thus appropriate to think of $R$ as a weighted Green's function. Since $P'(\omega)$ involves the imaginary part of $R(\omega)$, the photoluminescence spectrum at this mean-field level represents a weighted measure of the single particle density of states of the electron system with a localized hole.

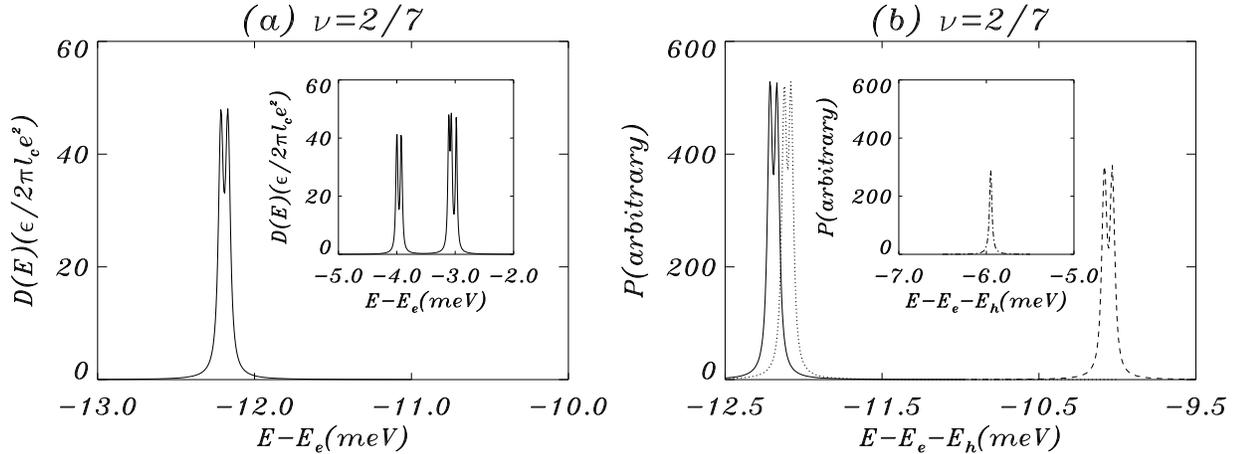

FIG. 1. Electron density of states and PL spectrum for perfect WC with no electron-hole interaction for $\nu = 2/7$. (a) Electron DOS below chemical potential at zero temperature (inset: DOS above chemical potential), where $E_e = \frac{1}{2}\omega_c$; (b) PL spectrum at different temperatures: $T = 0.0045 T_{melt}$(solid), $T = 0.45 T_{melt}$(dotted), $T = 0.9 T_{melt}$(dashed) and $T = 1.12 T_{melt}$(inset: dash-dotted).

The results of this mean-filed approach have discussed previously[7], so we only briefly summarize them here. The DOS for a perfect WC is a Hofstadter butterfly. For fractional filling factor $\nu = p/q$, this has $q$ subbands, and at zero temperature, the lowest $p$ bands are filled. We should expect to observe this structure; i.e., $p$ lines for $\nu = p/q$, in the photoluminescence spectrum of an ideal Wigner crystal if we turn off the electron-hole interaction. Presented in Fig.1 are the density of states (a) and the photoluminescence spectrum (b) for an undisturbed Wigner crystal at $\nu = 2/7$ with no e-h interaction. We can clearly see that the DOS has 7 bands, with only the lowest 2 occupied at low temperature. As expected, the photoluminescence has 2 peaks with a splitting identical to that of the DOS. An observation of this behavior in experiments would directly confirm the presence of a Wigner crystal in the system. Furthermore, as a function of temperature, the PL spectrum shifts very suddenly as the crystal melts to higher frequencies. This energy shift represents the latent heat of melting of the crystal. Once again, observation of this behavior would directly confirm the presence of a WC in the system.

While these effects are relatively simple to understand within the mean-field approach, one needs to question whether they can survive fluctuation effects. In particular, the smallness of the Hofstadter gaps in Fig. 1 suggest that they may not be robust enough to survive beyond the mean-field level. We will see for localized holes that this is indeed the case, and that this structure is replaced by a spectrum that reflects the collective mode density of states rather than the mean-field single particle spectrum. To show this, we proceed by reviewing how the collective modes are computed in the TDHFA, and then show how they may be incorporated into the PL spectrum.

### C. Collective Modes in Wigner Crystal

We define the dielectric function as the density response of a two-dimensional electron system to an external perturbation $U(\mathbf{p}, \tau)$, i.e.

$$\frac{\delta <\rho(-\mathbf{q}, \tau')>}{\delta U(\mathbf{p}, \tau)} \equiv -\chi(\mathbf{p}, \mathbf{q}; \tau - \tau'). \tag{31}$$

In the special circumstance of Wigner crystal, the density-density correlation function has the following special form:

$$\chi(\mathbf{G}_1 + \mathbf{q}, \mathbf{G}_2 + \mathbf{q}; \tau) = -g <T\tilde{\rho}(\mathbf{G}_1 + \mathbf{q}, \tau)\tilde{\rho}(-\mathbf{G}_2 - \mathbf{q}; 0)>, \tag{32}$$

where $\tilde{\rho} = \rho - <\rho>$. From now on, we denote $\chi(\mathbf{G}_1 + \mathbf{q}, \mathbf{G}_2 + \mathbf{q}; \tau)$ as $\chi_{\mathbf{G}_1 \mathbf{G}_2}(\mathbf{q}; \tau)$. The Fourier transform of this response function contains poles at the collective mode frequencies of the system. This function can be numerically computed using a generalized random phase approximation (GRPA)[12] which does not assume small displacements of the lattice electrons (as is necessary in classical approaches[9]), and thus gives a realistic dispersion relation for the collective modes across the entire Brillouin zone. In this subsection, we present the formalism for this GRPA calculation including the electron-hole interaction. (The magnetic length $l_c = \left(\hbar c/eB\right)^{1/2}$ will be set to unity).

In the real space, the dielectric function is given by:

$$\chi(X_1, X_2; X_3, X_4; \tau) = -g < T_\tau \tilde{\rho}(X_1, X_2)(\tau) \tilde{\rho}(X_3, X_4)(0) >, \tag{33}$$

where $\tilde{\rho}(X_1, X_2) = a^\dagger_{X_1} a_{X_2} - < a^\dagger_{X_1} a_{X_2} >$. The equation of motion for $\chi$ can be written as:

$$\begin{aligned}\frac{\partial \chi(X_1, X_2; X_3, X_4; \tau)}{\partial \tau} &= -g < [\tilde{\rho}(X_1, X_2), \tilde{\rho}(X_3, X_4)] > \delta(\tau) \\ &\quad -g < T_\tau [\mathcal{H} - \mu \hat{N}, \tilde{\rho}(X_1, X_2)](\tau) \tilde{\rho}(X_3, X_4)(0) > \\ &= -T_1 - T_2,\end{aligned} \tag{34}$$

where

$$\begin{aligned}T_1 &= g < [\tilde{\rho}(X_1, X_2), \tilde{\rho}(X_3, X_4)] > \delta(\tau) \\ &= g[< a^\dagger_{X_1} a_{X_4} > \delta_{X_2 X_3} - < a^\dagger_{X_3} a_{X_2} > \delta_{X_1 X_4}] \delta(\tau).\end{aligned} \tag{35}$$

We define the Fourier transform of this quantity as

$$\begin{aligned}T_1(\mathbf{q}, \mathbf{q}'; \tau) &= \frac{1}{g} \sum_{X_1 X_2} e^{-\frac{i}{2} q_x (X_1 + X_2)} \delta_{X_1, X_2 - q_y} \\ &\quad \times \frac{1}{g} \sum_{X_3 X_4} e^{-\frac{i}{2} q'_x (X_3 + X_4)} \delta_{X_3, X_4 + q'_y} T_1(X_1, X_2; X_3, X_4; \tau).\end{aligned} \tag{36}$$

so that in frequency and momentum space,

$$\begin{aligned}\tilde{T}_1(\mathbf{G} + \mathbf{q}, \mathbf{G}' + \mathbf{q}; \omega) &= <\rho(\mathbf{G} - \mathbf{G}')> \left[ e^{-i(\mathbf{G}+\mathbf{q}) \times (\mathbf{G}'+\mathbf{q})/2} \right. \\ &\quad \left. - e^{i(\mathbf{G}+\mathbf{q}) \times (\mathbf{G}'+\mathbf{q})/2} \right]\end{aligned} \tag{37}$$

The most important term $T_2$ comes from the electron-electron interaction and electron-hole interaction in the Hamiltonian. Following the method of Ref.12, a Hartree-Fock decomposition is applied to this term, to obtain

$$\begin{aligned}&\tilde{T}_2(\mathbf{G} + \mathbf{q}, \mathbf{G}' + \mathbf{q}; \tau) \\ &= 2i \sum_{\mathbf{G}''} \left[ W(\mathbf{G} - \mathbf{G}'') <\rho(\mathbf{G} - \mathbf{G}'')> + n_h V_{eh}(\mathbf{G} - \mathbf{G}'') e^{-(\mathbf{G} - \mathbf{G}'')^2/4} \right] \\ &\quad \times \sin[(\mathbf{G} + \mathbf{q}) \times (\mathbf{G}'' + \mathbf{q})/2] \chi_{\mathbf{G}'' \mathbf{G}'}(\mathbf{q}; \tau) \\ &\quad - 2i \sum_{\mathbf{G}''} W(\mathbf{q} + \mathbf{G}'') <\rho(\mathbf{G} - \mathbf{G}'')> \\ &\quad \times \sin[(\mathbf{G} + \mathbf{q}) \times (\mathbf{G}'' + \mathbf{q})/2] \chi_{\mathbf{G}'' \mathbf{G}'}(\mathbf{q}; \tau).\end{aligned} \tag{38}$$

Thus the equation of motion for $\chi_{\mathbf{G} \mathbf{G}'}(\mathbf{q})$ at specific wavevector $\mathbf{q}$ in the Brillouin zone can be expressed in terms of the following matrix equation:

$$\left[\omega + i\delta - A(\mathbf{q}) - D(\mathbf{q}) \tilde{W}(\mathbf{q})\right] \chi(\mathbf{q}) = D(\mathbf{q}), \tag{39}$$

where

$$A_{\mathbf{GG'}}(\mathbf{q}) = 2i \left[ W(\mathbf{G} - \mathbf{G'}) < \rho(\mathbf{G} - \mathbf{G'}) > + n_h V_{eh}(\mathbf{G} - \mathbf{G'}) e^{-(\mathbf{G}-\mathbf{G'})^2/4} \right]$$
$$\times \sin\left[(\mathbf{G} + \mathbf{q}) \times (\mathbf{G'} + \mathbf{q})/2\right]$$
$$D_{\mathbf{GG'}}(\mathbf{q}) = -2i < \rho(\mathbf{G} - \mathbf{G'}) > \sin\left[(\mathbf{G} + \mathbf{q}) \times (\mathbf{G'} + \mathbf{q})/2\right] \tag{40}$$
$$\tilde{W}_{\mathbf{GG'}}(\mathbf{q}) = W(\mathbf{q} + \mathbf{G})\delta_{\mathbf{GG'}}$$

The collective modes are given by the eigenvalues of the matrix $A + D\tilde{W}$. Note that generally this matrix is not Hermite. However, we can still diagonalize this matrix numerically and it turns out that all the eigenvalues are real (and symmetric around zero) for this matrix.

### D. Phonon Shake-up Theory

We now incorporate the effect of collective modes into the photoluminescence spectrum calculation. Recall that we need to calculate the quantity

$$\mathcal{R}_{ij}(X_1, X_2; i\omega_n) = -\int_0^\beta < T_\tau a_{X_1}(\tau) c_i(\tau) c_j^\dagger a_{X_2}^\dagger > e^{i\omega_n \tau} d\tau.$$

Working in the lowest Landau level, the equation of motion for $\mathcal{R}_{ij}(X_1, X_2; \tau)$ is given exactly by Eq.(25). In the derivation of the terms for contribution from the electron-electron interaction and the electron-hole interaction, by applying the many-body Hamiltonian given by Eq.(3), we encounter the following correlation function:

$$\mathcal{C}_{ij}(X_1 X_2; X_3 X_4; \tau', \tau) \equiv -g < T_\tau a_{X_1}^\dagger(\tau') a_{X_2}(\tau') a_{X_3}(\tau) c_i(\tau) c_j^\dagger(0) a_{X_4}^\dagger(0) >, \tag{41}$$

In the momentum space, this correlation function can be written as:

$$C_{ij}(\mathbf{q}, \mathbf{q}'; \tau', \tau) = \frac{1}{g} \sum_{X_1 X_2} e^{-\frac{i}{2}q_x(X_1+X_2)} \delta_{X_1, X_2 - q_y} \tag{42}$$
$$\times \frac{1}{g} \sum_{X_3 X_4} e^{-\frac{i}{2}q'_x(X_3+X_4)} \delta_{X_3, X_4 + q'_y} \mathcal{C}_{ij}(X_1, X_2; X_3, X_4; \tau', \tau).$$

Then the equation of motion for $R_{ij}(\mathbf{G}, \omega)$ may be written as

$$\frac{\partial}{\partial \tau} R_{ij}(\mathbf{G}, \tau) = < \rho(\mathbf{G}, \tau) > \delta_{ij}\delta(\tau) - \epsilon_0 R_{ij}(\mathbf{G}, \tau) \tag{43}$$
$$- n_h \sum_{\mathbf{G'}} V_{eh}(\mathbf{G'}) e^{i\mathbf{G'} \times \mathbf{G}/2 - G'^2/4} R_{ij}(\mathbf{G} - \mathbf{G'}, \tau)$$
$$- \frac{1}{S} \sum_{\mathbf{q} \neq 0} V_c(\mathbf{q}) C_{ij}(-\mathbf{q}, \mathbf{q} + \mathbf{G}) e^{i\mathbf{G} \times \mathbf{q}/2 - q^2/2}$$
$$- \frac{1}{S} \sum_{\mathbf{q}} V_{eh}(\mathbf{q}) C_{ij}(-\mathbf{q}, \mathbf{G}) e^{-i\mathbf{q} \cdot \mathbf{R}_i - q^2/4}.$$

In Eq.(43), $V_c(\mathbf{q})$ and $V_{eh}(\mathbf{q})$ are the Fourier transforms of the electron-electron and electron-hole interactions, respectively, $\epsilon_0$ is the energy of the localized hole, and the sum over $\mathbf{G'}$ is only over reciprocal lattice vectors, while the sums over $\mathbf{q}$ are over all wavevectors. $\mathbf{R}_i$ specifies the position of the hole in the $i$ th unit cell, and $< \rho(\mathbf{G}, \tau) > e^{-G^2/4}$ is the expectation value of a Fourier component of the electron density. While this quantity is independent of $\tau$ in the groundstate, it will be convenient for later purposes to formally leave it as an argument of the density. The method for computing these Fourier components has been described previously[7,12]. Eq.(43) represents the first in an infinite series of equations relating an $n$ particle Green's function to the $n + 1$ particle Green's function[15]. In the mean-field approximation, it was simplified by employing a Hartree-Fock (HF) decomposition of Eq.(41), which converts Eq.(43) into a self-consistent equation for $R_{ij}$[7].

To include shakeup effects, we instead extend this hierarchy to one more level, writing down a self-consistent form for $\mathcal{C}_{ij}$ which explicitly contains the collective mode excitations. To carry out this program, it is convenient to implicitly define a self-energy by writing the last two terms in Eq.(43) as

$$-\sum_{\mathbf{G'}} \int_0^\beta d\tau' \Sigma(\mathbf{G}, \mathbf{G'}; \tau - \tau') R_{ij}(\mathbf{G'}, \tau'),$$

which implicitly defines the self-energy for this problem. The HF approximation for the PL is equivalent to replacing $\Sigma(\mathbf{G}, \mathbf{G}'; \tau - \tau')$ by $\Sigma^{HF}(\mathbf{G}, \mathbf{G}')\delta(\tau - \tau')$, where

$$\Sigma^{HF}(\mathbf{G}, \mathbf{G}') = W(\mathbf{G} - \mathbf{G}') <\rho(\mathbf{G} - \mathbf{G}'; \tau)> e^{i\mathbf{G}\times\mathbf{G}'/2} \qquad (44)$$

$$+ \frac{1}{2\pi} \sum_{\mathbf{q}} V_{eh}(\mathbf{q}) <\rho(-\mathbf{q}, \tau)> e^{-q^2/4 - i\mathbf{q}\cdot\mathbf{R}_i} \delta_{\mathbf{G},\mathbf{G}'}, \qquad (45)$$

and $W$ is the sum of the direct and exchange Coulomb potentials[7,12]. (Note here we keep the negative sign in front of $\mathbf{q}$ because it is not necessarily a reciprocal lattice vector.) Thus we need to calculate the self-energy

$$\Sigma = \Sigma^{HF}\delta(\tau - \tau') + \delta\Sigma$$

beyond mean-field approximation.

To generate a self-consistent equation for $\mathcal{C}_{ij}$, we take a functional derivative[15] of Eq.(43) with respect to an external potential $U$ which contributes an extra term in the Hamiltonian

$$\mathcal{H}_{ext} = \sum_{X_1 X_2} U(X_1, X_2) a^{\dagger}_{X_1} a_{X_2}.$$

In doing this, it must be noted that a term directly coupling the Green's function $R_{ij}$ to the external potential $U$

$$-\sum_{\mathbf{p}} R_{ij}(\mathbf{p}, \tau) U(\mathbf{G} - \mathbf{p}, \tau) e^{\frac{i}{2}\mathbf{G}\times\mathbf{p}}$$

must be added to Eq.(43), and the sums over reciprocal lattice vectors must be extended to all wavevectors, because $U$ is not in general commensurate with the lattice.

We introduce a generalized Green's function $F$ satisfying

$$R_{ij}(\mathbf{p}, \tau) \equiv \sum_{\mathbf{p}'} F(\mathbf{p}, \mathbf{p}', \tau) <\rho(\mathbf{p}')> . \qquad (46)$$

The equation of motion for the generalized Green's function can be obtained following Eq.(43)

$$\frac{\partial}{\partial \tau} F(\mathbf{p}, \mathbf{p}_1, \tau - \tau_1) = \delta_{\mathbf{p}\mathbf{p}_1}\delta(\tau - \tau_1) - \epsilon_0 F(\mathbf{p}, \mathbf{p}_1, \tau - \tau_1) \qquad (47)$$

$$-n_h \sum_{\mathbf{q}} V_{eh}(\mathbf{q}) e^{i\mathbf{q}\times\mathbf{p}/2 - q^2/4} F(\mathbf{p} - \mathbf{q}, \mathbf{p}_1, \tau - \tau_1)$$

$$- \sum_{\mathbf{q}} \int_0^{\beta} d\tilde{\tau} \Sigma(\mathbf{p}, \mathbf{q}, \tau - \tilde{\tau}) F(\mathbf{q}, \mathbf{p}_1, \tilde{\tau} - \tau_1), \qquad (48)$$

Under Hartree-Fock approximation, the equation of motion can be simplified as

$$\frac{\partial}{\partial \tau} F^{HF}(\mathbf{p}, \mathbf{p}_1, \tau - \tau_1) = \delta_{\mathbf{p}\mathbf{p}_1}\delta(\tau - \tau_1) - \epsilon_0 F^{HF}(\mathbf{p}, \mathbf{p}_1, \tau - \tau_1) \qquad (49)$$

$$-n_h \sum_{\mathbf{q}} V_{eh}(\mathbf{q}) e^{i\mathbf{q}\times\mathbf{p}/2 - q^2/4} F^{HF}(\mathbf{p} - \mathbf{q}, \mathbf{p}_1, \tau - \tau_1)$$

$$- \sum_{\mathbf{q}} \Sigma^{HF}(\mathbf{p}, \mathbf{q}) F^{HF}(\mathbf{q}, \mathbf{p}_1, \tau - \tau_1),$$

Comparing Eq.(49) with the equation of motion for the single particle Green's function (Eqs.(12) and (13)), one can derive the following relation:

$$F^{HF}(\mathbf{p}_1, \mathbf{p}_2; \tau_1 - \tau_2) = -e^{(E_0 - \omega_0 - \mu_e)(\tau_1 - \tau_2) + i\mathbf{p}_1\times\mathbf{p}_2/2} G(\mathbf{p}_1 - \mathbf{p}_2; \tau_1 - \tau_2). \qquad (50)$$

Here $\mathbf{p}_1 - \mathbf{p}_2$ has to be one of the reciprocal lattice vectors. Using these definitions, we find Eqs. (43) and (48) may be rewritten as

$$F(\mathbf{p}_1,\mathbf{p}_2;\tau_1-\tau_2) = F^{HF}(\mathbf{p}_1,\mathbf{p}_2;\tau_1-\tau_2) \qquad (51)$$
$$-\sum_{\mathbf{q}_1\mathbf{q}_2}\int\int d\tilde{\tau}_1 d\tilde{\tau}_2 F^{HF}(\mathbf{p}_1,\mathbf{q}_1;\tau_1-\tilde{\tau}_1)$$
$$\times \delta\Sigma(\mathbf{q}_1,\mathbf{q}_2;\tilde{\tau}_1-\tilde{\tau}_2) F(\mathbf{q}_2,\mathbf{p}_2;\tilde{\tau}_2-\tau_2)$$

and

$$R_{ij}(\mathbf{G},\tau) = R_{ij}^{HF}(\mathbf{G},\tau) - \sum_{\mathbf{G}_1\mathbf{G}_2}\int\int d\tilde{\tau}_1 d\tilde{\tau}_2 F^{HF}(\mathbf{G},\mathbf{G}_1;\tau-\tilde{\tau}_1) \qquad (52)$$
$$\times \delta\Sigma(\mathbf{G}_1,\mathbf{G}_2;\tilde{\tau}_1-\tilde{\tau}_2) R_{ij}(\mathbf{G}_2,\tilde{\tau}_2).$$

Finally, we define the zeroth order Green's functions $R_{ij}^0$ and $F^0$ which satisfy Eqs. (48) and (46) with the self-energy $\Sigma$ (and external potential) set to zero. It is not difficult to show that

$$R_{ij}(\mathbf{q},\tau) = R_{ij}^0(\mathbf{q},\tau) - \sum_{\mathbf{q}_1\mathbf{q}_2}\int\int d\tilde{\tau}_1 d\tilde{\tau}_2 F^0(\mathbf{q},\mathbf{q}_1;\tau-\tilde{\tau}_1) \qquad (53)$$
$$\times \Sigma(\mathbf{q}_1,\mathbf{q}_2;\tilde{\tau}_1-\tilde{\tau}_2) R_{ij}(\mathbf{q}_2,\tilde{\tau}_2).$$
$$-\sum_{\mathbf{q}_1\mathbf{q}_2}\int d\tilde{\tau} F^0(\mathbf{q},\mathbf{q}_1;\tau-\tilde{\tau})$$
$$\times U(\mathbf{q}_1-\mathbf{q}_2;\tilde{\tau})e^{i\mathbf{q}_1\times\mathbf{q}_2/2}R_{ij}(\mathbf{q}_2,\tilde{\tau}). \qquad (54)$$

It is this form of the equation of motion for $R_{ij}$ that is most convenient for taking a functional derivative. Using standard methods[15], we find the Green's functions $R_{ij}$ and $C_{ij}$ satisfy the relation

$$\frac{\delta R_{ij}(\mathbf{p},\tau)}{\delta U(\mathbf{p}',\tau')} = C_{ij}(\mathbf{p}',\mathbf{p};\tau',\tau) - gR_{ij}(\mathbf{p},\tau)<\rho(\mathbf{p}')>_{\tau'}$$
$$= \delta C_{ij}(\mathbf{p}',\mathbf{p};\tau',\tau). \qquad (55)$$

Taking the functional derivative of Eq. (54), and defining $C_{ij}^{HF}$ as the Hartree-Fock decomposition of $C_{ij}$ that turns Eq. (43) into Eq. (27), we arrive after much algebra at the relation

$$\delta C_{ij} = \delta C_{ij}^{HF} + C_{ij}^{PH},$$

where

$$C_{ij}^{PH}(\mathbf{p}',\mathbf{p};\tau',\tau) = -\sum_{\mathbf{q}_1\mathbf{q}_2}\int d\tilde{\tau}_1 d\tilde{\tau}_2 F^{HF}(\mathbf{p},\mathbf{q}_1;\tau-\tilde{\tau}_1)$$
$$\times \frac{\delta\Sigma(\mathbf{q}_1,\mathbf{q}_2)_{\tilde{\tau}_1-\tilde{\tau}_2}}{\delta U(\mathbf{p}',\tau')} R_{ij}(\mathbf{q}_2,\tilde{\tau}_2). \qquad (56)$$

$C_{ij}^{PH}$ is the correction to $C_{ij}$ beyond mean-field theory, and it will contain the phonon (i.e., collective mode) contribution to $C_{ij}$.

To this point we have made no approximations. To obtain a closed expression for $C_{ij}^{PH}$, we need an approximate form for $\Sigma$. The best choice available is the Hartree-Fock form, Eq. (45), and so we make this substitution. Combining Eqs.(31) and (44), we get:

$$\frac{\delta\Sigma^{HF}(\mathbf{q}_1,\mathbf{q}_2)_{\tilde{\tau}}}{\delta U(\mathbf{p}',\tau')}$$
$$= -W(\mathbf{q}_1-\mathbf{q}_2)e^{i\mathbf{q}_1\times\mathbf{q}_2/2}\chi(\mathbf{p}',\mathbf{q}_2-\mathbf{q}_1;\tau'-\tilde{\tau})$$
$$-\frac{1}{2\pi}\sum_{\mathbf{q}}V_{eh}(\mathbf{q})e^{-q^2/4-i\mathbf{q}\cdot\mathbf{R}_i}\delta_{\mathbf{q}_1\mathbf{q}_2}\chi(\mathbf{p}',\mathbf{q};\tau'-\tilde{\tau}). \qquad (57)$$

Upon substituting the resulting $C_{ij}^{PH}$ into the following equation (the phonon mode contribution to Eq.(43))

$$\sum_{\mathbf{G}_2} \int d\tau_2 \delta\Sigma(\mathbf{G}_1, \mathbf{G}_2; \tau_1 - \tau_2) R_{ij}(\mathbf{G}_2, \tau_2)$$

$$= \frac{1}{2g} \sum_{\mathbf{q}} W(\mathbf{q}) e^{i\mathbf{G}_1 \times \mathbf{q}/2} C_{ij}^{PH}(-\mathbf{q}, \mathbf{G}_1 + \mathbf{q}; \tau_1)$$

$$+ \frac{1}{S} \sum_{\mathbf{q}} V_{eh}(\mathbf{q}) e^{-q^2/4 - i\mathbf{q}\cdot\mathbf{R}_i} C_{ij}^{PH}(-\mathbf{q}, \mathbf{G}_1; \tau_1), \qquad (58)$$

one obtains the self-energy contributed by collective modes:

$$\delta\Sigma(\mathbf{G}_1, \mathbf{G}_2, \tau_1 - \tau_2)$$
$$= \frac{1}{4\pi} \sum_{\mathbf{G}_1' \mathbf{G}_2'} \int_{BZ} d^2\mathbf{q} \, e^{i\mathbf{q}\times(\mathbf{G}_1 - \mathbf{G}_2)/2 - i\mathbf{G}_1 \times \mathbf{G}_1'/2 + i\mathbf{G}_2 \times \mathbf{G}_2'/2}$$
$$\times W(\mathbf{G}_1' + \mathbf{q}) W(\mathbf{G}_2' + \mathbf{q})$$
$$\times F^{HF}(\mathbf{G}_1 - \mathbf{G}_1' - \mathbf{q}, \mathbf{G}_2 - \mathbf{G}_2' - \mathbf{q}; \tau_1 - \tau_2) \chi_{\mathbf{G}_1' \mathbf{G}_2'}(\mathbf{q}; \tau_1 - \tau_2)$$
$$+ \frac{1}{(2\pi)^3} \sum_{\mathbf{G}_1' \mathbf{G}_2'} \int_{BZ} d^2\mathbf{q} \, e^{-(\mathbf{G}_1' + \mathbf{q})^2/4 - (\mathbf{G}_2' + \mathbf{q})^2/4}$$
$$\times V_{eh}(\mathbf{G}_1' + \mathbf{q}) V_{eh}(\mathbf{G}_2' + \mathbf{q})$$
$$\times F^{HF}(\mathbf{G}_1, \mathbf{G}_2; \tau_1 - \tau_2) \chi_{\mathbf{G}_1' \mathbf{G}_2'}(\mathbf{q}; \tau_1 - \tau_2). \qquad (59)$$

In Eq.(59), $\int_{BZ} d^2\mathbf{q}$ represents an integral over wavevectors in the first Brillouin zone of the superlattice. With this expression, we are now able to compute the PL intensity. We substitute $\Sigma = \Sigma^{HF} \delta(\tau - \tau') + \delta\Sigma$ into Eq.(43), and Fourier transform this with respect to imaginary time. This means that a Fourier transform of Eq.(59) will be necessary, leading to frequency summations of the form (suppressing wavevector arguments) $\sum_{i\omega_n} F(i\omega_n) \chi(\mathbf{q}; i\omega - i\omega_n)$. To accomplish this, we represent $\chi$ as a sum over its collective mode poles[12]; the frequency sums may be then computed using standard methods[15]. The computation of $\delta\Sigma$ is clearly the bottleneck in this computation, since it requires two reciprocal lattice sums and an approximate sum over the continuous wavevector $\mathbf{q}$. We have accomplished this using 469 $\mathbf{q}$ points in the first Brillouin zone, for one and three electrons per unit cell, which give very similar results. Finally, once we have computed $\delta\Sigma$, it is straightforward to substitute this into the frequency version of Eq.(43), obtain $R(\mathbf{G}, \omega)$, and from there compute the PL spectrum.

### III. NUMERICAL RESULTS

Following the algorithm in the proceeding section, one can calculate the collective modes distribution (density of states) and their effects on the photoluminescence spectrum beyond the mean-field limit. In this section, we present our numerical results for the cases with different strength of electron-hole interaction and different filling factors. We will also discuss the temperature dependence of the phonon shakeup effect in photoluminescence from strong magnetic field induced Wigner crystal.

#### A. General Result

Shown in Fig.2(b) is our calculated photoluminescence spectrum for filling fraction $\nu = 1/5$ at $T = 0$ (electron density $N_s = 6 \times 10^{10} cm^{-2}$). Our hole is assumed to be strongly localized, and located $250 \mathring{A}$ from the electron plane. We ignore the electron-hole interaction in this case. A well-defined shakeup peak may be seen approximately 2 meV below the main PL peak; a second very weak satellite is observed approximately 3.5meV below the main peak. The origins of these peaks may be understood in terms of the phonon DOS, which is illustrated in Fig.2(a). A van Hove singularity, arising from zone-edge phonons, appears as a strong double peak near 0.4meV. Two other peaks may be seen near 1.2 meV and 1.9 meV. There are weak sidebands associated with each of these peaks in the PL spectrum. The precise interpretation of these peaks is unclear; however, it has been speculated that these represent vacancy-interstitial excitations[12]. We point out that it is *crucial* to use a fully quantum mechanical treatment of the collective excitations of the lattice to observe these higher order satellites; classical treatments of the phonons[9,11] do not produce these unusual excitations.

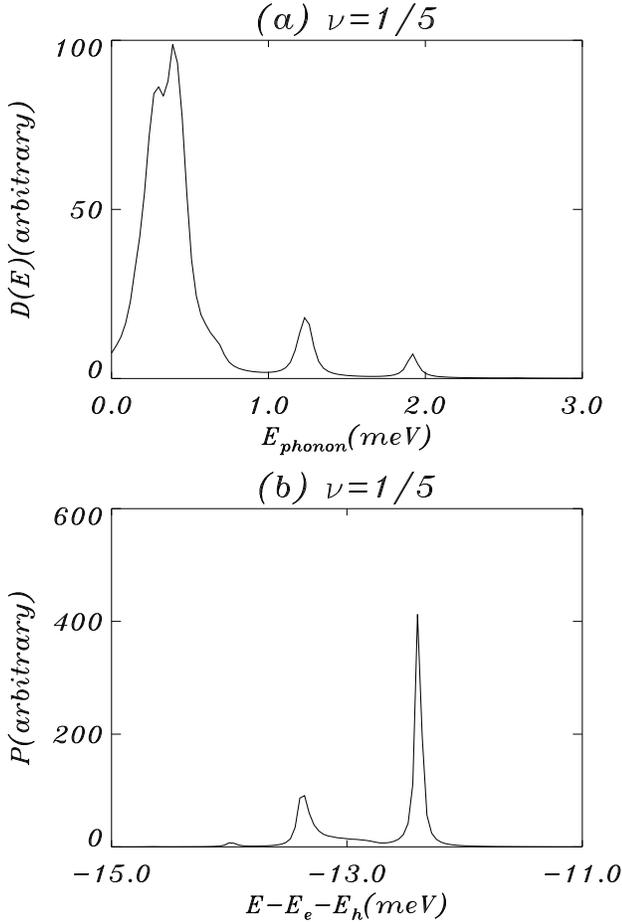

FIG. 2. Phonon density of states(a) and shakeup spectrum(b) for a perfect WC with no electron-hole interaction for $\nu = 1/5$.

It should also be noted that the splitting between the main PL peak and the first sideband is actually larger than the energy at which the van Hove singularity in the phonon DOS appears. The reason for this is that there is a strong self-energy renormalization due to the phonons in the main PL peak. Physically, this arises because the final state of the crystal contains a vacancy, which is lowered in energy by a distortion of the lattice – i.e., by allowing the electrons surrounding the vacancy to relax inward. The self-energy shift accounts for this lowering in energy of the final state of the WC, and leads to an upward shift of the main PL peak. Our calculation shows that the final states in which phonons are excited are not nearly so strongly renormalized by lattice relaxation effects, leading to the increased splitting between the main PL peak and the sidebands.

### B. Effect of Electron-hole Interaction

Presented in Fig.3 are the corresponding phonon density of states (a) and shake-up spectrum in the presence of strong electron-hole interaction for $\nu = 1/5$ at $T = 0$. Since in this case, we include a finite number of electrons in one supercell to account the lattice distortion due to the attraction between the localized hole and electrons, there is more structure in the phonon density states as shown in Fig.3(a).

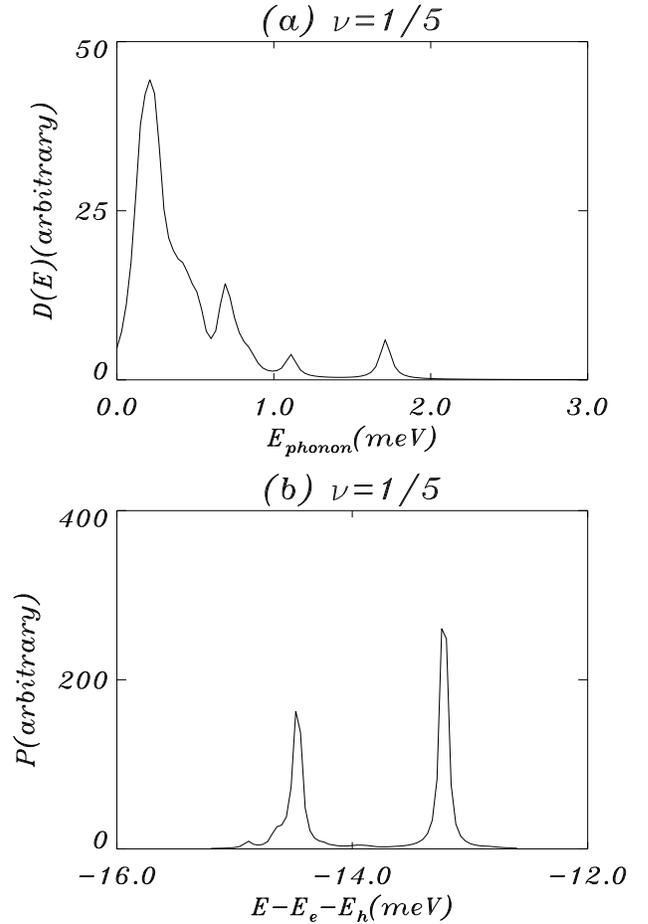

FIG. 3. Phonon density of states(a) and shakeup spectrum(b) for a WC in the prresence of electron-hole interaction for $\nu = 1/5$. There are 3 electrons per unit cell.

We also notice that the oscillator strength of the shakeup peak increases significantly from the case with no electron-hole interaction. Since zone-edge phonons correspond to short-range collective modes near the center of the supercell where the hole is located, strong electron-hole interaction does enhance their contribution to the shakeup spectrum. Some of these local collective modes even persist above the melting temperature because of the non-uniform electron density distribution near the hole.

### C. Other Filling Factors

Fig.4 illustrates our results for $\nu = 2/7$ in the absence of electron-hole interactions. We can see splittings in the phonon density of states in Fig.4(a) which are re-lated to the splittings in the Hofstadter spectrum. However, zone-edge phonon modes still dominate the shakeup spectrum, and as can be seen in Fig.4(b), except for a change in energy scale caused by changing the magnetic field, the photoluminescence lineshape is essentially identical to the case of $\nu = 1/5$. This contrasts sharply with the results found in the mean-field approximation[7], where without electron-hole interactions, a filling $\nu = p/q$ generally yields $p$ distinct lines for a localized hole (see Fig.1). While the splittings are so small in that situation that they are difficult in practice to resolve, evidently shakeup effects wipe out this structure even in principle.

be able to identify this characteristic spectrum of bands and gaps that is unique to a WC in a magnetic field in itinerant hole experiments.

### D. Temperature Dependence

Fig.5 illustrate the evolution of the photoluminescence spectrum with increasing temperature in the absence of electron-hole interaction. As shown in the figure, in the solid phase, the main peak slightly shifts to higher energy with increasing temperature as in the mean-field case. It undergoes a sudden shift upon melting and stays at higher energy in the liquid phase. In contrast, the shakeup sidebands slightly shift to lower energy with increasing temperature in the solid phase widening the gap between the main peak and the sidebands. However the sidebands also shifts rapidly back to higher energy upon melting and disappear in the liquid phase. This is necessarily so, because in the melted phase, the density is uniform without electron-hole interaction, and there are no collective modes in the lowest Landau level[12]. In this situation, phonon satellites uniquely distinguish between a liquid and a solid state.

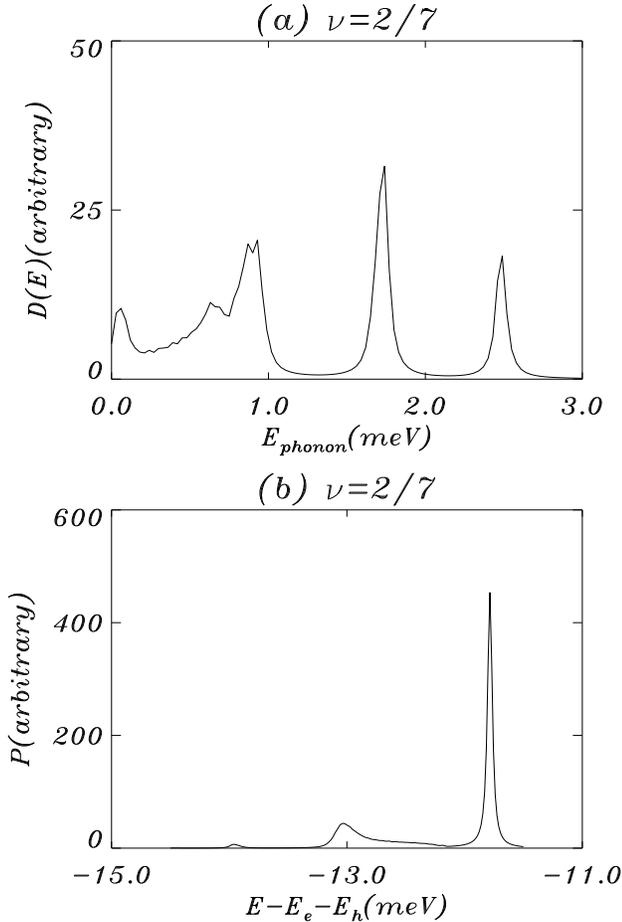

FIG. 4. Phonon density of states(a) and shakeup spectrum(b) for a perfect WC with no electron-hole interaction for $\nu = 2/7$.

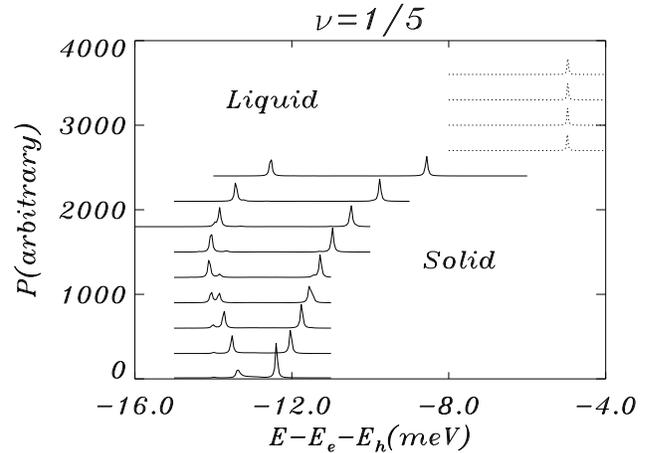

FIG. 5. Temperature dependence of the PL shakeup spectrum for a perfect WC with no electron-hole interaction for $\nu = 1/5$. Starting from the lowest curve, the corresponding temperatures comparing to the melting temperature are $T/T_{melt}$ =0.004, 0.112, 0.224, 0.336, 0.448, 0.561, 0.673, 0.785, 0.897, 1.009, 1.121, 1.233, 1.345.

We note that for the case of an itinerant hole, we expect these characteristic splittings to survive shakeup effects. The reason is that (neglecting excitonic effects), the PL spectrum is related to the product of Green's functions for the hole and the electron. For localized holes, only the latter has poles in the form of a Hofstadter spectrum at the mean-field level, which are wiped out by shakeup effects. However, for itinerant holes, the hole Green's function also has poles of the Hofstadter form, which are unaffected by shakeup. Thus, one should in principle

Fig.6 illustrate the temperature dependence of the photoluminescence spectrum in the presence of strong electron-hole interaction. In the solid phase, one can see that the qualitative behavior of the main peak and the sidebands are the same as in the absence of electron-hole interaction. However, upon transition, the phonon shakeup sidebands undergo a sudden shift to higher energy and still persist in the liquid state. In this case, since there is a non-uniform electron density near the hole, allowing some local collective modes to persist even above the melting temperature. Except the sudden shifts of the phonon sidebands upon melting, they could not provide an definitive evidence for Wigner crystal. However, with further increase in temperature, or increased setback between the hole and the 2DEG (weakening the electron-hole interaction), the oscillator strength of sidebands in the liquid phase will significantly decrease. Observation of this behavior *would* distinquish the liquid from the solid phase.

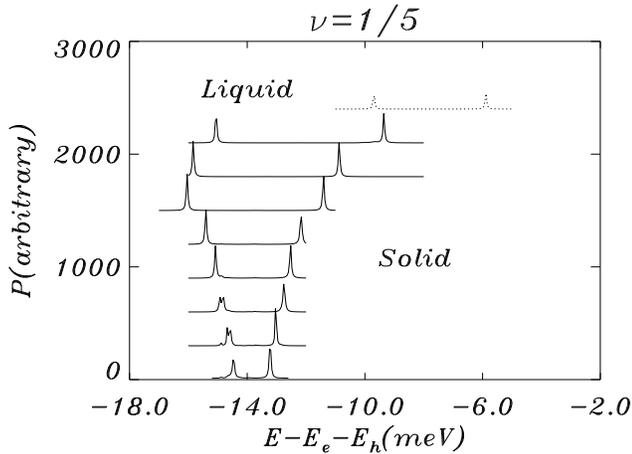

FIG. 6. Temperature dependence of the PL shakeup spectrum for a WC in the presence of electron-hole interaction for $\nu = 1/5$. Starting from the lowest curve, the corresponding temperatures comparing to the melting temperature are $T/T_{melt}$ =0.004, 0.045, 0.090, 0.134, 0.224, 0.448, 0.673, 0.897, 1.121.

Our results near the melting temperature should be viewed with caution, and taken as a qualitative rather than a quantitative picture. This is because the melting temperature is still calculated within the mean-field Hartree-Fock approximation, in which only the contribution from particle-hole pair excitations is included, and thermally excited collective modes are neglected. The calculated melting temperature is thus significantly higher than the experimentally observed transition temperature (about $1.5K$). However, our results should be more reliable at low temperatures since the collective mode contribution is expected to be insignificant there.

## IV. CONCLUSIONS

In summary, we have developed a method by which shakeup effects in the PL spectrum of a WC from localized holes may be computed, that treats the tunneling electron and the lattice electrons on an equal footing, and uses a fully quantum treatment of the collective modes that is realistic over the entire Brillouin zone. Our method is quite general, and should be applicable to other shakeup problems where quantum fluctuations are important. We find that the Hofstadter spectrum found in a mean-field analysis of this experiment is lost (although we expect it to survive in itinerant hole experiments), and is replaced by a series of sidebands due to creation of phonons and other collective excitations of the WC. These sidebands are a unique signature of the WC which are associated with zone-edge phonons, and can in principle be used to distinguish between a liquid and crystal state of the electrons. We find that there is a sudden shift in the PL spectrum upon melting of the crystal and disappearance of the sidebands in the liquid state in the case with weak electron-hole interaction. However, in the presence of strong electron-hole interaction sidebands persist above melting temperature.

## ACKNOWLEDGMENTS

The authors thank Dr. René Côté for helpful discussions. This work was supported by the NSF, through Grant Nos. DMR 95-03814 and DMR 91-23577, and by the US-ONR. HAF is supported by the Alfred P. Sloan Foundation and by a Cottrell Scholar Award of Research Corporation.